# Reconstruction of metabolic networks from high-throughput metabolite profiling data: *in silico* analysis of red blood cell metabolism

Ilya Nemenman[1,2], G. Sean Escola[1,5], William S. Hlavacek[2,3], Pat J. Unkefer[4], Clifford J. Unkefer[4], Michael E. Wall[1,2,4]

[1]Computer, Computational and Statistical Sciences Division
[2]Center for Nonlinear Studies
[3]Theoretical Division
[4]Bioscience Division
Los Alamos National Laboratory, Los Alamos, NM 87545
[5]Columbia University Medical Center, New York, NY10032

Corresponding authors:
1. Ilya Nemenman, CCS-3, MS-B256, Los Alamos National Laboratory, Los Alamos, NM 8545. Phone: 505-665-8250. Fax: 505-667-1126. E-mail: nemenman@lanl.gov
2. Michael E. Wall, CCS-3, MS-B256, Los Alamos National Laboratory, Los Alamos, NM 8545. Phone: 505-665-4209. Fax: 505-667-1126. E-mail: mewall@lanl.gov

## Abstract
We investigate the ability of algorithms developed for reverse engineering of transcriptional regulatory networks to reconstruct metabolic networks from high-throughput metabolite profiling data. For this, we generate synthetic metabolic profiles for benchmarking purposes based on a well-established model for red blood cell metabolism. A variety of data sets is generated, accounting for different properties of real metabolic networks, such as experimental noise, metabolite correlations, and temporal dynamics. These data sets are made available online. We apply ARACNE, a mainstream transcriptional networks reverse engineering algorithm, to these data sets and observe performance comparable to that obtained in the transcriptional domain, for which the algorithm was originally designed.

## 1. Introduction: The need for Benchmark Data

In the recent years, high-throughput (HTP) microarray profiling has generated large data sets that characterize the simultaneous activities of, essentially, all genes in a cell. These data sets have been used successfully to reverse engineer (RE) cellular transcriptional regulatory networks (see, for example, [1-3] for a collection of references). Similar experimental progress is expected in the emerging field of metabolomics, where sensitive HTP measurements of (relative or absolute) concentrations of many metabolites in a sample of cells are now possible in different preparations under various experimental interventions, and/or steady state growth conditions [4-6]. Anticipating the resulting data sets, there is a strong interest in development of computational tools that, unlike more traditional approaches based on sequence information [7] or chemical reactivity and conservation laws [8, 9], would use the relevant HTP data to expand our knowledge of metabolic networks, which, as extensive as it is, is still incomplete. Because metabolic networks share features with transcriptional regulatory networks, it is tempting to transfer successful methods developed in the context of transcriptional networks, such as those in [1-3, 10], to inference of metabolic networks.

An obvious advantage of transferring these methods is that only minimal modifications are required to the very extensive RE code base. On the other hand, it is not obvious that the existing methods will perform well on metabolic networks. Indeed, despite the superficial similarity, metabolic and transcriptional networks are quite different. In the transcriptional case, a transcription factor (TF), a parent, causes a change in the expression of its target gene, a child, without any direct effects on its own activity. This leads to correlations among expressions of TFs and their targets, and these can be readily discovered by various statistical techniques. Conversely, in metabolism, a substrate (a parent) is transformed into a product (a child). Thus, an increase in the child's abundance comes at the cost of decreasing the abundance of the parent. We therefore expect that the statistical associations in metabolic data will differ from those in gene expression data sets in unknown ways. Furthermore, the experimental noise has a tendency to mask interactions of low-mean or low-variance species. This has been a problem even in transcriptional analysis (e.g., spurious interactions in the ribosomal complex in [1]), where the expression levels and the involved characteristic time scales of reaching steady states are largely uniform across all genes. On the other hand, kinetic rates in a metabolic network can vary over many orders of magnitude for different species. Thus the time required for an organism to achieve a metabolic steady state can vary from milliseconds to hundreds of hours [11]. Furthermore, many metabolites are short-lived and low-abundance, and a "fully expressed" metabolite can mean anywhere from a few molecules to a few million molecules per cell, making consideration of the measurement noise very important.

Because of these differences between transcription and metabolism, the fidelity of standard transcriptional RE algorithms for metabolic networks cannot be assumed. It is therefore useful to test these methods on benchmark data that resemble real metabolic measurements, and for which the ground truth structure of the network is known. We are unaware of the existence of experimental data sets of this kind, and therefore we turn to numerical simulations. However, existing synthetic data sets have focused on realistic modeling of transcriptional regulation [12], and they may not represent metabolism well. Therefore, in this work, we undertake the task of generating synthetic benchmark metabolic data by using a well-established kinetic model of red blood cell (RBC) metabolism [11], which involves 39 metabolites connected by 44 individual reactions. These data have been made publicly available at http://www.menem.com/~ilya/wiki/index.php/RBC_Metabolic_Network. We then use ARACNE, a modern transcriptional network RE algorithm, which was developed and validated for gene expression analysis [10], to infer metabolic interactions from these synthetic metabolic data, and we argue that its performance is comparable to that in the transcriptional case. This outcome suggests that other transcriptional HTP-based RE algorithms might be transferred to the domain of metabolism with minimal changes as well.

## 2. The RBC Metabolic Benchmark Data

In generating synthetic benchmark data, our goal is not to accurately simulate a real system. Rather, our goal is to exercise transcriptional RE algorithms by generating data that are complex enough to incorporate different features of metabolism (dynamic ranges, temporal properties, correlations among chemical species, noises, etc.), but are still simple enough to analyze in detail. Specifically, we generated four data sets to account for ever more complex scenarios of realistic profiling of RBC metabolism (see below). As the majority of transcriptional RE methods take steady-state abundance data as inputs, we focused on steady state metabolic profiling in three of these synthetic datasets, and studied dynamics only in the fourth one.

To generate the simulated data, we modified a publicly available Mathematica workbook implementation of the RBC model [11]. The model has 5 parameters that can be controlled externally: the Donnan ratio, $R$, (which determines the difference in the pH inside and outside of the cell); glucose concentration, $G$; total intracellular magnesium concentration, both free and bound, $Mg$; intracellular inorganic phosphorus concentration, $Pi$; and extracellular (plasma) sodium concentration, $Na$. For the first three data sets, these external control parameters were sampled at random 1000 times from specified probability distributions, representing different experimental setups, and the steady state values of the metabolic network were found by using the methods in the RBC workbook. In a significant number of situations (up to 30% or more depending on the data set), the randomly selected parameters did not lead to steady state solutions. These samples were removed from the data set.

<u>Data set 1 (chemostat):</u> This data set simulates RBC steady-state measurements from chemostat culture experiments. All the parameters are uncorrelated, uniformly distributed variables, with the ranges indicated below (the numbers in parentheses are the values of the parameters assumed in the RBC workbook model [11]). The ranges were established by a literature search for conditions of various culture experiments that did not lead to an immediate cell death. We emphasize again that our aim is not to accurately model specific biochemical experiments—instead, our aim is to provide test data with realistic features. Hence the crude specification of the parameter ranges below.
1. $R = 0.2…1.6$ (0.6). The natural value of $R$ seems to be hard to pinpoint [13, 14] (see also discussion of Data set 2), but experiments on prepared/perturbed cultures achieve $R$ as high as 1.6 [13, 15]. The lowest value of 0.69 comes from [13], which is higher than the value of 0.6, used in [11]. However, [16] suggests that the internal-to-external $Pi$ concentration ratio (which is closely related to $R$) can go down to 0.2 for pH near 8.0. We chose this value as the lower limit on $R$, even though it is probably too low in the context of the RBC (pH=7.4). For $R > 0.8$, the RBC dynamical system often does not have accessible steady-state solutions (depending on the other control parameters).

2. $G$ = 2.0…30.0 mM (5.0) [17].
3. $Mg$ = 0.1…20 mM (2.7) [15]. For larger values of $Mg$, steady states are hard to find, and we do not include such parameter combinations in the data set.
4. $Pi$ = 0.6…1.8 mM (1.2) [16].
5. $Na$ = 100…200 mM (140). Identifying this parameter from culture experiments is difficult, since most data are about internal, rather than plasma sodium. However, Refs. [18, 19] note clinical cases with $Na$ down to ~110 and up to ~180, in which the patient still survived. In view of this, the range of 100-200 mM for culture experiments seems reasonable.
Additionally, we observed that an external pH of 7.55 is normal for culture conditions, and values down to pH=7.0 [15] and up to 8.0 [16] have been recorded.

Data set 2 (natural): This data set represents the variability of RBC metabolite concentrations in blood samples from healthy humans. The control parameters are taken as uncorrelated normal variables with means $\mu_i$ and standard deviations $\sigma_i$ (indicated as $\mu_i \pm \sigma_i$ below), where $i = 1…5$ denotes the identity of the parameter. We take physiologically allowable intervals found in the literature as $\pm 2\sigma$ around the mean.
1. $R$ = 0.75 ± 0.1. Ref. [14] gives .825 for normal human. The RBC model [11] uses 0.6, citing [13], which suggests $R$=0.69. At the normal pH=7.4, [15] suggests that the internal-to-external $Pi$ ratio (and hence $R$) is between 0.4 and 1.0, with the median about 0.8. Given cell preparations for all of these analyses, neither of the values may be anatomically relevant, and the real value is likely unknown for in vivo conditions. Hence we've chosen the distribution for which the mean is roughly the average of the reported human data, and the range of the normal data is about $\pm 2\sigma$ around the mean.
2. $G$ = 5 ± 0.6 mM [20, 21].
3. $Mg$ = 3.3 ± 0.2 mM [15, 22].
4. $Pi$ = 0.9 ± 0.15 mM. This estimate is based on $Pi$ values between 0.6 and 1.2 [16], obtained for an external pH of 7.4 (the default value of the RBC model). Values reported in alternative sources (1.0 in [23], 0.8 in [24], and 0.98 in [25]) differ from these by less than $2\sigma$. It is important to realize that the RBC model uses intracellular concentration as the control parameter, while most references, such as [20, 22], focus on plasma concentrations, leading to large discrepancies.
5. $Na$ = 140 ± 2.5 mM [20, 22]. Values reported in alternative sources differ from these by less than $2\sigma$.
Additionally, the following information was collected: normal external pH of 7.24 [15] or 7.4 [16] for an unperturbed cell.

Data set 3 (correlated): This data set attempts to model the *in vivo* metabolite concentrations more faithfully by incorporating physiological correlations among the controls. Using this data set, one may study effects of the correlations (and thus reduction of the dynamic ranges of the response variables) on the performance of RE algorithms. For most of the parameters, we were unable to find quantitative measurements of the correlation coefficients in the literature, and instead only trends were reported. We summarized the trends into correlation coefficients $\rho_{ij}$ of 0 (no trend, or no data available), $\pm 0.3$ (weak correlation), and $\pm 0.5$ (strong correlation). Then the data set was generated by sampling the control parameters from multivariate normal distributions with means and variances as in Data Set 2, and with the correlation coefficients summarized in Table 1.

1. *Pi-R* and *Mg-R*: The *Pi*-pH and *R*-pH correlation coefficients are -0.85 and -0.6…-0.76 (for different species) respectively [16]. Thus, it is reasonable to assume that the *Pi-R* correlation is positive and large (+0.5 in our notation). Also notice that the Donnan ratio should have similar correlations with all internal ionic concentrations (modulo the sign of the charge). An agreement with the *Mg-R* value [15] is encouraging.

2. *Na-R*: Recall that *Na* is an extracellular concentration and the correlation with *R* is not obvious. Nonetheless, [26], eq. 2, suggests a negative correlation. This, however, may be affected by fluctuations of the total sodium level and of cell volumes. Therefore, we choose a value of -0.3 for this correlation.

3. *Na-G*: A small positive correlation is reported in [26].

4. *G-R*: While we found no explicit data relevant for estimating this correlation, *G* is positively correlated with *Mg* and *Na* ions, which are, in turn, negatively correlated with *R* (see above). Thus, a small negative value for the *G-R* correlation is assumed.

5. *Na-Pi*: Ref. [25] suggests strong positive pair-wise correlations between the internal *Na* and *Na* efflux, between the internal *Pi* and the inverse of the *Na* efflux, and between the external and internal *Na*. Overall, we deduce a weak negative correlation between external *Na* and internal *Pi*; this correlation is further supported by the opposite sign charges of these particles.

6. *G-Pi*: Significant negative correlation is reported in [27].

7. *Na-Mg*: Weak competitive behavior between these species is reported in [28].

Data set 4 (evolving parameters): The RBC model takes up to 100 hours or more to reach a steady state [11]. However, in a natural environment, the control parameters fluctuate on time scales less than an hour. In fact, it takes only tens of seconds for blood to circulate. Further, the same drop of blood visits the liver every 20 minutes or so, and this may completely change the concentration of various ions in the cells and around them. While we found no explicit data about the temporal variability of the five control parameters in humans, we believe it is reasonable to model each of them as correlated Ornstein-Uhlenbeck processes with the means, the standard deviations, and the species-species correlations as in Data set 3, and the correlation time $\tau = 20$ min for each process. This data set required the most extensive changes to the RBC model, enabling dynamic variation of the control parameters during the temporal evolution of the system. The resulting time series data represent 20 hours of evolution of the RBC model, sampled every 10 seconds (for a total of 7201 samples); researchers may subsample the series and/or shorten it to better match the sampling scheme of an actual experiment. This data set is designed to test the application of network reconstruction in the presence of non-equilibrated, time-series data.

Table 1. Correlation coefficients of the five control parameters of the RBC model. References used to set the values are listed as well.

|   | R | G | Mg | Pi | Na |
|---|---|---|----|----|-----|
| R |   | -0.3 (see text) | -0.5 [15] | +0.5 [15, 16] | -0.3 [26] |
| G |   |   | +0.3 [29] | -0.5 [27] | +0.3 [26] |

| | | | | | 0 | -0.3 [28] |
|---|---|---|---|---|---|---|
| *Mg* | | | | | | |
| *Pi* | | | | | | -0.3 [25] |
| *Na* | | | | | | |

Noise: We simulate experimental errors in each metabolite concentration by adding Gaussian zero-mean random noise to the output of RBC model. The noise variance is given by $A^2 + B^2 x^2$, where $x$ is the abundance, and $A$ and $B$ describe the contribution of the absolute and the relative noise components. Each of the four simulated data sets is available from our web site with a multitude of $A$ and $B$ values. This model is a good noise model for transcriptional data [30], and we expect it to be relevant for metabolism as well. However, specialized noise studies will have to be performed on real experimental data, once available, to verify this model. Notice, in particular, that this model does not take into the account errors that may emerge due to the overlap of peaks in mass-spectrometry based metabolite profiling.

## 3. Reverse Engineering of Metabolic networks with ARACNE algorithm

With availability of the data secured, we tested whether transcriptional RE tools, exemplified by the ARACNE algorithm [10], can be used with minimal modifications for analysis of metabolic networks.

Like many other network reconstruction methods, ARACNE models dependencies among activity variables (e.g., gene expressions or metabolite concentrations), $\{g_i\}$, as probabilistic, defined by an unknown probability distribution $P(\{g_i\})$. A probabilistic description reflects the effects of unobserved molecular species and of experimental noise. A bona fide biological interaction corresponds to a nonzero statistical dependency between activity profiles, measured by the mutual information $I(g_i, g_j) = \langle \log P(g_i, g_j) / P(g_i) P(g_j) \rangle$. Evaluating the mutual information and identifying its value above a certain threshold with an interaction is the basis of the Relevance Networks (RN) method [31]. However, in a major problem for most RE algorithms, as signals propagate through the networks, many non-interacting species also become correlated and result in a positive $I$ (e.g., two non-interacting targets of the same transcription factor may be highly statistically dependent). To isolate statistical interactions that have the highest chance to correspond to real biological interactions, ARACNE then uses the data processing inequality (DPI) [10] after statistically significant values of the mutual information have been identified. The DPI states that, if stochastic variables $g_1$ and $g_3$ interact only through a third one, $g_2$, then $I(g_1, g_3) \leq \min\left[I(g_1, g_2); I(g_2, g_3)\right]$. Thus ARACNE analyzes each gene triplet and designates the link with the lowest MI value as indirect. To minimize the effect of incorrect estimations of the MI, this designation is made only when a link is at least $\tau$ % below the second weakest one. For $\tau \sim 5\ldots15\%$, the method has been validated in synthetic [10] and in vitro [1] transcriptional networks.

To establish a metric for the fidelity of a reconstruction, we note that the RBC interaction network is specified by a system of first-order differential equations

$dx_i/dt = f(\{x_{j \in Ne(i)}\})$, where $Ne(i)$ is a set of neighbors of the node $i$, including all immediate parents, children, and modulators (effectors) of the reactions. Thus, the steady state probability distribution is $P(i\,|\,\text{rest of network}) = P(i\,|\,Ne(i))$, which corresponds to linking metabolites in the interaction graph to all of their neighbors [10]. This results in a "gold standard" adjacency matrix with 107 pairwise interactions among metabolites, to which the ARACNE reconstruction is to be compared.

As a first check, we reconstructed the RBC metabolic network using just 19 conditions from Data Set 1 with $\tau = 0$ and with no modifications of ARACNE's parameters from their transcription-tuned values, and using the algorithm implementation reported in Ref. [32]. A total of 14 interactions are predicted, 11 of which are substantiated by the model; this is a recall of 10% (14 out of 107) at a precision of 78%. (11 out of 14)). We then performed a systematic study using Data Set 1 with added measurement noises of different levels, modeling real experiments. Since the noise properties are different from those in the transcriptional case, we expended a substantial effort to fine-tune the internal ARACNE parameter essential for estimation of mutual information, the *kernel width* [10, 32], for each run. When an observed metabolite variance across different steady states becomes smaller than the associated experimental noise variance, establishing its interactions is impossible by any statistical method, thus we remove such metabolites from the network, and all indirect interactions mediated by them are considered direct for the validation purposes. For a small and dense network, like the RBC one, where 14% percent of all metabolite pairs are connected by interactions, the node removal sets a limit on realistic values of the noise: at high noise variance, the network becomes a small and almost fully connected cluster, making precision of the algorithm artificially high.

Precision vs. Recall Curves (PRCs) for noisy and noiseless data and for two different DPI tolerance values corresponding to the Relevance Networks and the ARACNE algorithms are shown in Figure 1. These curves are generated by adjusting the significance threshold for mutual information estimation so that metabolite pairs with a mutual information below the threshold are not allowed to participate in an interaction. Higher thresholds decrease the number of putative interactions, which eliminates most of the false positives, thus increasing the precision. Lower thresholds admit more pairs for consideration, and lead to higher recalls. For small, dense, and loopy RBC network, we expect the precision to drop fairly fast as the recall increases, and this is clearly seen in the Figure. Further, since precisions of 15-20% correspond to a "by chance" performance, only the top left corner of the Figure shows any improvement over the random assignment of interaction edges. Still, the most significant feature in this figure is that, for both noise levels, and for both ARACNE and RNs, there is always a range with a non-negligible recall and with precision ~1. Thus the algorithms can be tuned to produce a (small) number of predictions that are highly likely to be true. Furthermore, in the relevant region of high precision, the low tolerance (ARACNE) lines are substantially higher than the high tolerance (RN) ones. Just like in the case of transcription [10], this indicates an improvement from using ARACNE over RNs on metabolic data. Thus, a minimally modified ARACNE algorithm can be used to accurately predict metabolic interactions.

In Figure 2, we compare how the change in the dynamic range of responses due to smaller variability and correlations in the external control parameters affects the validity of the reconstruction. These effects can be observed only for large noises, which are larger than the signal for some metabolites. Hence we used the noise setting such that the effective number of nodes in the *chemostat* dataset is 19. Surprisingly, while this number is smaller in the *natural* dataset, corresponding to the smaller response variability, it is larger again in the *correlated* dataset. This holds true for other noise levels and means that parameter correlations are synergistic in their effect on variability of responses. The *natural* dataset PRC seems to indicate the best performance. However, comparison to the other PRCs is not fair since the effective network is smaller and the "by chance" precision (35%) is higher in this case. On the other hand, the *chemostat* network has a smaller chance precision (32%) than the *correlated* one (34%), yet its PRC is higher. This indicates that the decreased response variance and/or spurious correlations among metabolites introduced by correlations among control parameters indeed decrease the quality of the reconstruction.

Finally, in Figure 3 we examine applicability of ARACNE to time-dependent metabolic data. Specifically, we would like to understand how temporal correlations among subsequent samples affect reconstruction. In order not to confound dynamics with the data set size, here we always reconstruct the network from 400 samples. However, in different runs, these samples are spaced every 10, 40, and 160 seconds apart, with the temporal correlations among subsequent samples dropping in proportion to the dilution. For comparison, we also plot reconstruction based on 400 independently sampled steady states from Data Set 3. We clearly see that, while reconstruction with temporally-correlated data is possible, the corresponding PRCs are generally lower than for the steady-state case and drop dramatically faster. This indicates that methods that explicitly consider temporal structure [3, 33] should be used instead of steady-state methods, like ARACNE. Surprisingly, the steady-state PRC is lower than its counterparts for some of the temporal data for very large precision values. While this may be an insignificant artifact, it can also mean that weak, yet fast, metabolite interactions unobservable in steady state may become evident in a dynamic setting, when their effect is not masked by strong, slow interactions. Finally, a low starting point of the 10 s sampling curve may be a result of the noise masking small abundance changes between subsequent samples at fine temporal discretization.

## 4. Conclusions

We have generated benchmark synthetic metabolic data sets and analyzed them with ARACNE, a representative transcriptional networks reverse engineering algorithm. The performance of the algorithm for metabolic networks is comparable to that for transcriptional ones. This finding may be considered as a basis for an optimistic view that transcriptional networks RE algorithms may be transferred *en mass* with relatively few modifications to metabolic applications. However, this ease of transfer must be verified on a case-by-case basis. Importantly, we now have synthetic benchmarking data sets, which can aid in this verification. The most important limitation of the data sets is the relatively small size of the RBC metabolic network, which limits the ability to ascertain statistical significance of many findings. For example, due to this problem, we cannot verify whether the MINDY algorithm [34], which is an extension of ARACNE capable of elucidating interactions that are statistically insignificant overall, but become apparent in subsets of data, provides an advantage over ARACNE proper when applied to this metabolic data.

The steady-state values in data sets 1-3 are ideal for application of ARACNE, RN and other steady-state algorithms. However, because it may take many hours for a controlled culture to equilibrate, time-resolved assays might enable more rapid inference of metabolic interactions. Data set 4 provides a means for testing metabolic network inference using such experiments, and can potentially aid in the design of efficient sampling schemes that minimize the cost of reconstructing metabolic networks.

Finally, we notice that performance of transcriptional RE algorithms on metabolic data can be substantially improved beyond that observed in Figs. 1-3. Indeed, not every biochemical reaction is possible in nature. For example, metabolic reactions conserve mass, and, unlike in transcription, this places strict limits on which species can be metabolically coupled. Similarly, atomic species are also conserved by metabolism, which places even more constraints on allowable reactions, akin to [8]. Metabolic profiling almost always involves detailed determination of metabolite masses, and frequently provides information about the chemical structure of the compounds using isotopic labeling. It is, therefore, essential to incorporate these constraints into HTP profiling-based methods, such as ARACNE, in the future.

## Acknowledgements
This work was supported by DOE/NNSA under the contract No. DE-AC52-06NA25396. IN and WSH were further supported by NIH award 1 R21 GM080216 01.

# Figures

**Figure 1.**

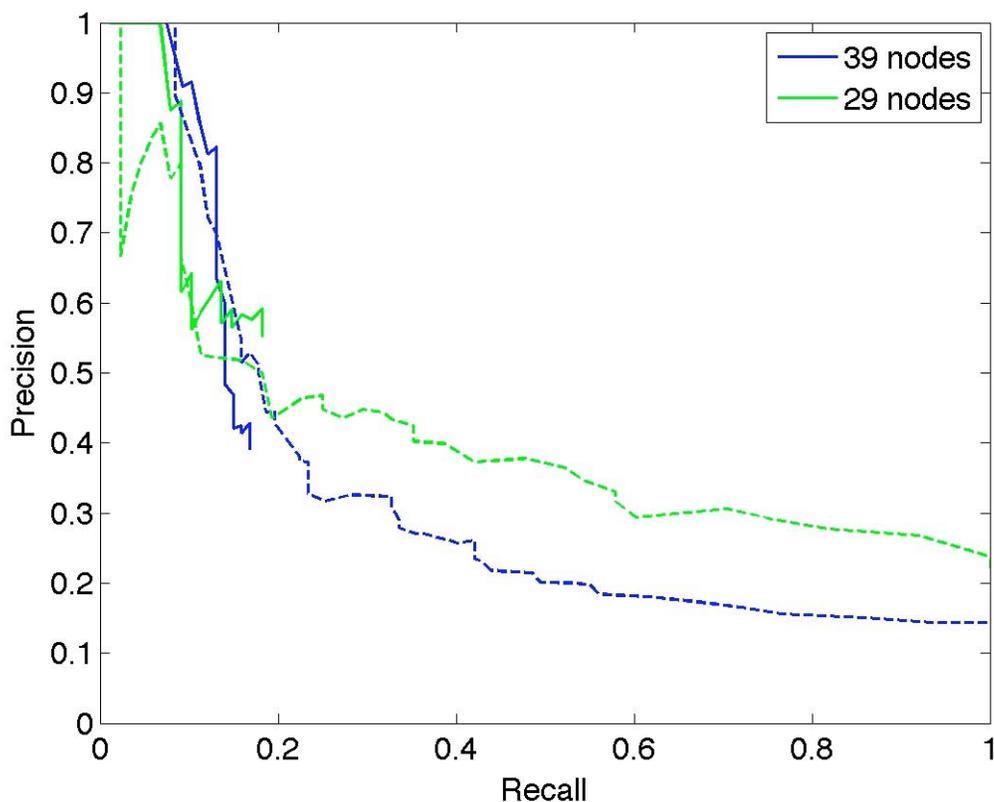

Precision vs. Recall curves for different noise levels (blue – no noise, green – $A = 0.001$, $B = 0$), indexed by the number of nodes with the variance above the experimental noise (39 and 29 respectively). Dashed lines were constructed with the DPI tolerance of 100% (no DPI applied, equivalent to the Relevance Networks algorithm), and solid lines have 0% DPI tolerance (pure ARACNE algorithm). Curves corresponding to other values of the tolerance generally fall between these two extremes. At recall of 1.0, the precision is 14% and 22% for blue/green lines, respectively, and it corresponds to 107/88 true positives out of 741/406 possible metabolite pairs, all indexed as putative interactions if no DPI is applied, and if all mutual information values are treated as significant.

**Figure 2.**

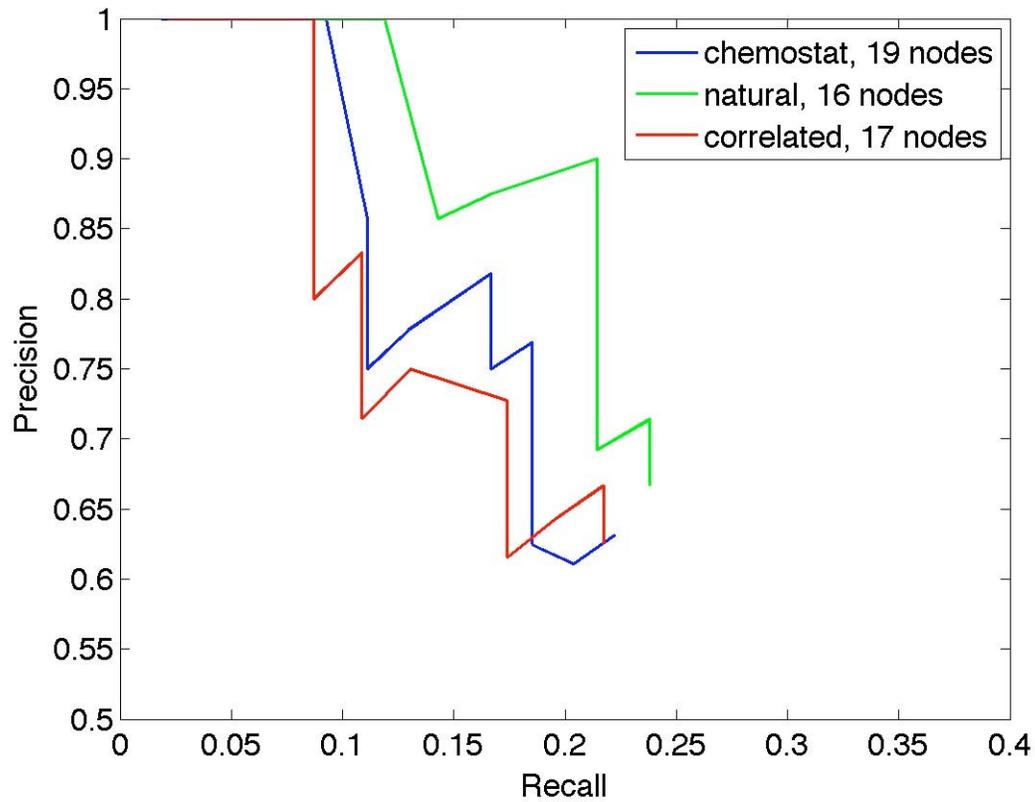

Precision vs. recall curves for data sets with different distributions of the control parameters and the same noise ($A = 0.01$, $B = 0$). For the three data sets, the number of true interactions (chance precision) is 32%, 35%, and 34% of the total number of possible metabolite pairs. Note unconventional scaling of the axes.

**Figure 3.**

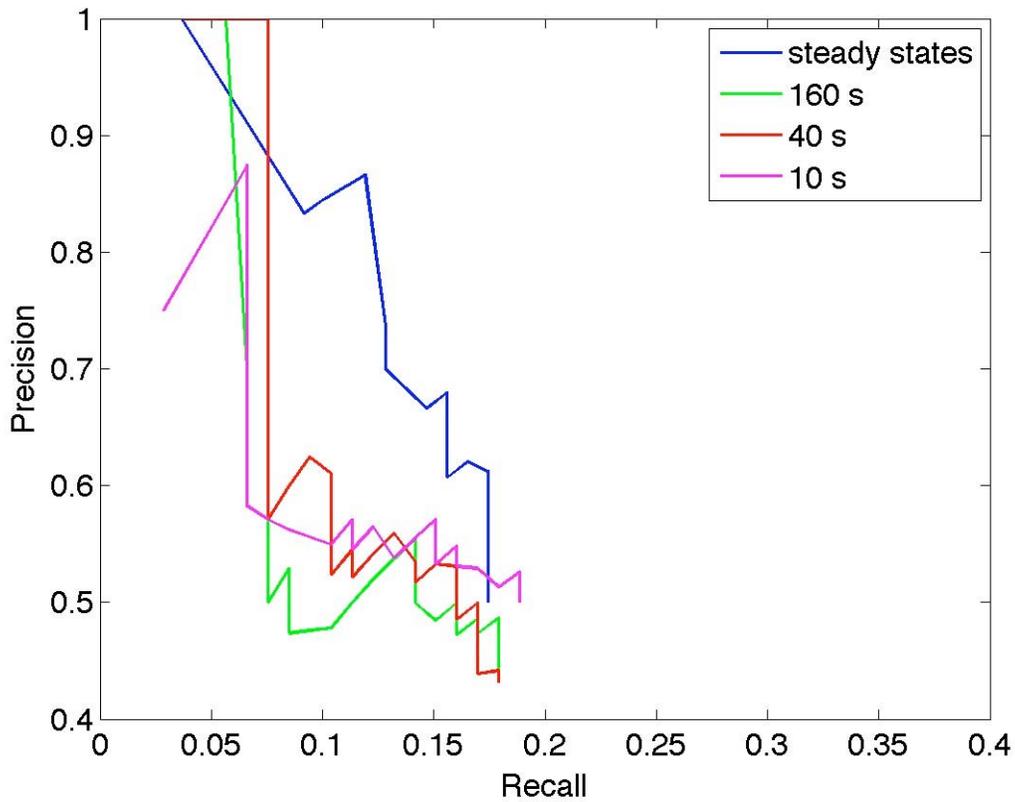

Precision vs. recall curves for dynamical data using 400 samples sampled at different intervals. Steady state reconstruction with the same number of samples is shown for comparison. For these plots, the effective number of nodes is 37, 38, 38, and 38 respectively. Note unconventional scaling of the axes.